Constructivism in Online Science Courses

Marta R. Stoeckel

Department of Curriculum & Instruction, University of Minnesota

CI 8572: Learning Theory and Classical Research in STEM Education

May 11, 2020

**Abstract**

Reform movements in science education, such as inquiry-based instruction, have been heavily influenced by constructivist learning theories (National Research Council, 2000). These learning theories place the learner as the sole constructor of knowledge and emphasize the importance of the learner's inquiry process (Yilmaz, 2008). In constructivist inquiry-based science education, lab experiences frequently play an important role in instruction as they provide students with opportunities to observe and make sense of the world around them (National Research Council, 2000), which raises the question of how inquiry-based science instruction can be translated to online environments. There are several models for lab experiences in online science courses, including hands-on labs where students directly manipulate materials, remote labs where students manipulate materials through a computer, and virtual labs and simulations where students work with simulated materials (Powell, et al., 2010). Hands-on labs play an important role, especially given constructivist views that students construct meaning by making observations of the world around them, but there is evidence that simulated and virtual labs can play an important role and may even be better suited to some instructional goals than hands-on labs. Constructivist instruction also requires students to make their process visible and teachers to be responsive to student thinking, both of which are more challenging in online environments (Crippen, et al., 2013). However, with intentional design, these features can be incorporated into online science courses (Jaber, et al., 2018; Jang, 2009).

## Constructivism on Online Science Courses

As of 2019, at least 310,000 K-12 students were enrolled in fully online schools, with many more taking a mix of online and face-to-face courses (Evergreen Education Group Digital Learning Collaborative, 2019). This number has been steadily growing, creating a need to examine effective pedagogies for online environments. In science education in particular, where current calls for reform are heavily influenced by constructivist learning theories, there is a need to determine how to design online learning environments that enable students to participate in the process of knowledge construction. In this literature review, I will begin by examining the influence of constructivist learning theories in current science education reform movements, especially inquiry-based instruction. Next, I will examine the challenges of implementing constructivist science instruction in online environments, with particular attention to the challenge of providing meaningful lab work in an online course. Finally, I will describe some examples of constructivist science instruction in online courses.

### Constructivist Learning Theories

Constructivism describes a range of learning theories built on the premise that knowledge cannot be transmitted; rather, it is constructed by individuals as they make sense of observations and experiences (von Glasersfeld, 1983). All of these theories frame learning as an active process where the teacher acts as a guide and facilitator as learners explore and organize information, rather than as an expert responsible for dispensing "truth" to students. While all constructivist learning theories place the individual as the sole constructor of knowledge, different constructivist learning theories account for different contextual factors and take varying stances on the nature of knowledge (Yilmaz, 2008).

von Glasersfeld (1983) proposed radical constructivism, which argues that as learners construct their understanding, they are also constructing their own reality. Therefore, there are no independently existing objective facts; there are only the ways an individual experiences and understands the world around them. This reinforces the view of the teacher's role as guide since if there is no objective truth for the teacher to dispense, the teacher can only support students finding useful exploration and developing their individual truth.

Bruner (1961), by contrast, emphasizes the role of discovery in learning, which implies an underlying truth available to be discovered. Bruner describes discovery as any experience where an individual obtains new knowledge for themselves through their own efforts, even if that knowledge has already been obtained by others. In addition to granting learners space to construct knowledge, Bruner sees the opportunity to engage in the process of discovery as an important learning experience in and of itself. He suggests that the process of discovery has associated techniques and heuristics. These skills cannot be taught; they can only be obtained through practice and experience with the process of discovery.

Both perspectives on knowledge construction emphasize the importance of the learner engaging in reflection. von Glasersfeld (1983) suggests that reflection on action is part of how the learner constructs their reality, while Bruner (1961) sees reflection as part of how a learner refines their skills in inquiry and discovery. Most critically, as with any constructivist learning theory, both agree that knowledge is always constructed by the individual, not transmitted from an external source.

**Constructivism in Science Education Reform**

The influence of constructivist learning theories on science education is most apparent in calls for inquiry-based instruction. Inquiry-based instruction refers to instructional practices that engage students in the process of science in order to develop new knowledge (National Research Council, 2012), an approach especially influenced by Bruner's call for learners to engage in the process of discovery in order to construct new knowledge. The National Research Council (2000, p. 25) identifies five essential features of inquiry for science classrooms:

1. Learners are engaged by scientifically oriented questions.

2. Learners give priority to evidence, which allows them to develop and evaluate explanations that address scientifically oriented questions.

3. Learners formulate explanations from evidence to address scientifically oriented questions.

4. Learners evaluate their understandings in light of alternative explanations, particularly those reflecting scientific understanding.

5. Learners communicate and justify their proposed explanations.

These essential features reflect a view of learning that prioritizes students' active construction of new knowledge over the transmission of facts.

The National Research Council (2012) also takes the stance that developing students' skills in the process of inquiry is of equal importance in science education as the science content itself. This view aligns with Bruner's (1961) perspective that the techniques, heuristics, and skills of discovery are an important outcome of learning through discovery. This view is reflected in the Next Generation Science Standards (NGSS Lead States, 2013) through the inclusion of eight science and engineering practices that students should engage in during science instruction:

1. Asking questions and defining problems

2. Developing and using models

3. Planning and carrying out investigations

4. Analyzing and interpreting data

5. Using mathematics and computational thinking

6. Constructing explanations and designing solutions

7. Engaging in argument from evidence

8. Obtaining, evaluating, and communicating information

The National Research Council (2012) emphasizes that students should engage in these practices in order to construct an understanding of the targeted science content.

Inquiry-based instruction is often associated with an emphasis on lab experiences as opportunities for students to develop knowledge (National Research Council, 2000). This aligns with the constructivist view that a learner constructs knowledge through observation, experience, and experimentation in the world around them. It is important to note that the presence of lab experiences alone do not guarantee an inquiry-based or constructivist learning experience. In K-12 classrooms, labs are frequently teacher-directed with rigid procedures provided to students. In addition, labs in K-12 classrooms are often used to verify concepts provided by the teacher or a textbook, rather than as an opportunity for students to construct knowledge. Labs in K-12 classrooms are also often disconnected from other elements of instruction, placing them as isolated experiences rather than an integral part of knowledge development (National Research Council, 2006).

In order for lab experiences to more closely reflect the essential features of inquiry-based instruction, the National Research Council (2006, p. 3) suggests seven learning goals for lab experiences in inquiry-based classrooms:

1. Enhancing mastery of subject matter

2. Developing scientific reasoning

3. Understanding the complexity and ambiguity of empirical work

4. Developing practical skills

5. Understanding the nature of science

6. Cultivating interest in science and interest in learning science

7. Developing teamwork abilities

These goals call for open-ended labs that allow students to become active participants in the process of discovery, as described by Bruner (1961). The National Research Council (2006) also calls for lab experiences to be interwoven with other aspects of classroom instruction in order to place labs as a component of knowledge construction, rather than a disconnected experience.

Meeting these calls for truly constructivist, inquiry-based instruction requires what has been termed responsive instruction (Jaber, et al., 2018). During responsive instruction, the teacher solicits the questions, ideas, and other input from students, then designs instruction based on disciplinary connections to that input. By making students' thinking explicit and using that thinking to design instruction, the teacher is able to emphasize students' sense-making and knowledge construction in the classroom.

While lab experiences are an important part of inquiry-based science instruction, they do not inherently reflect constructivist learning theories in science classrooms. However, when used in a responsive classroom that places student thinking at the foreground of instruction to provide students with the opportunity to construct knowledge and when structured to emphasize the process of inquiry, they become a crucial component of an inquiry-based, constructivist science classroom.

## Lab Activities in Online Science Instruction

The growing prevalence of online instruction raises the question of how current movements promoting constructivist, inquiry-based science instruction can be applied to online environments. In particular, given the crucial role of lab experiences in inquiry-based science instruction, how can online instruction provide students with meaningful inquiry-based lab experiences?

Powell, et al. (2010) define four types of labs that can be conducted in online science courses: hands-on, virtual, simulated, and remote. In hands-on labs, students directly manipulate physical equipment, as is typically envisioned for science labs. For online courses, these labs will typically either utilize equipment likely available to students at home or provide students with a lab kit that includes the required materials. Virtual labs allow students to set up virtual equipment and manipulate variables. Simulated labs also utilize virtual equipment, but only allow students to manipulate variables, not to set up the equipment, narrowing the scope of possible experiments within the lab. Remote labs rely on physical equipment, but allow students to manipulate the materials remotely. Similar to simulated labs, students typically have limited ability to modify the set up, restricting the breadth of possible experiments. While these are framed as distinct categories, Ma and Nickerson (2006) point out that the boundaries are often blurry, especially as technology is increasingly integrated into hands-on labs. In particular, virtual and simulated labs are often placed into a single category by both practitioners and researchers (Crippen, et al., 2012).

In a position statement on the role of lab investigations in science instruction, the National Science Teaching Association (2007) takes the stance that while virtual and simulated labs may provide instructional value, they are not suitable for lab investigations. This is consistent with the constructivist view that it is vital for students to directly experience, observe, and experiment on the physical world in order to engage in the active construction of meaning and develop scientific knowledge (Jang, 2009). Even when simulated and virtual labs are based on actual data, Srinivasan, et al. (2006) found students perceive simulated and virtual labs as disconnected from the physical world, which may limit the degree to which students connect their constructed knowledge to reality. Lindsey and Good (2005) found that students who used simulated and virtual labs also had less awareness and understanding of the context of an experiment than students who conducted remote or hands-on labs. By contrast, Ma and Nickerson (2006) suggest that regardless whether a lab is hands-on, remote, virtual, or simulated, the key factor in the outcomes of a lab experience is how students' perceive the lab's presence. If students perceive the lab as connected to the real, physical world, it has the potential to engage students in the sense making required for knowledge construction. If students see the lab as disconnected from the real world, even a hands-on lab is unlikely to engage students in meaningful sense making and knowledge construction.

There is little empirical evidence to support a preference for hands-on labs over virtual, simulated, or remote labs. In a literature review, Brinson (2015) found most empirical studies showed online labs resulted in equal or greater gains in content knowledge than hands-on labs. In addition, while few empirical studies attempted to measure gains in students' inquiry skills, the studies that did examine inquiry skills found no meaningful difference between online and in-person labs. Similarly, in a study on students' conceptual understanding of heat and temperature, Zacharia and Olympiou (2011) found students who used a simulation had similar conceptual understanding as students who conducted a hands-on lab. The key factor seemed to be that students had the opportunity to manipulate variables and engage in the inquiry process, not whether students had the opportunity to physically interact with equipment.

Authentic experiments in the form of hands-on and remote labs provide students with important opportunities to develop skills in the process of inquiry that are difficult to replicate in simulated or virtual labs. In particular, authentic experiments naturally include experimental uncertainties, systematic errors, and other ambiguities and messiness that complicate data analysis and cannot be truly replicated by simulated or virtual labs (Tho & Yeung, 2018). This provides students with the opportunity to learn to navigate these challenges during data analysis to find meaning in imperfect data sets. Interestingly, Lindsay and Good (2005) found that students engaged in a remote lab were more likely to recognize nonidealities in their data than peers engaged in a comparable hands-on lab. The remote lab required less attention to manipulate equipment than the hands-on lab, which allowed students to attend more carefully to interpreting their data.

Simulated and virtual labs may provide students with more opportunity to engage in metacognition during lab experiences. Lindsay and Good (2005) found that students who conducted a simulation lab were more able to focus on the process of learning, including asking and answering their own questions and reflecting on their thinking, than students who completed an equivalent remote or hands-on lab. Since knowledge construction requires students to engage in reflection on their thinking and process (Bruner, 1961; von Glasersfeld, 1983), this suggests a potential benefit for students using a simulated lab.

Hands-on labs may be preferable when developing teamwork abilities is a goal of the lab. Corter, et al. (2011) found that when students conducted a hands-on lab, they formed a group identity early in the process which carried over into the data analysis and interpretation phases of the lab. By contrast, groups that used a remote lab or a simulation did not begin to form a group identity until much later in the lab and had significantly less cohesion as a result. Students were able to complete the data collection for the remote and simulated labs asynchronously and individually, while the hands-on lab required students to collaborate during data collection, which likely led to the increased cohesion.

Ultimately, the most crucial factor for achieving inquiry-based lab experiences in an online science course is not whether the lab is hands-on, remote, virtual or simulated, but how the lab is used in the course of instruction. While many online lab experiences are not constructivist, many in person lab experiences are not constructivist, either (Brinson, 2015). If the lab is used responsively to foreground student thinking, allows students to engage in science practices and the process of inquiry, aligned to the identified instructional goals, and is designed for students to construct, rather than verify new knowledge, it can serve a constructivist classroom, regardless of the form it takes.

### Making Process Visible

One of the biggest challenges in online inquiry-based science instruction is the ways teacher-student interactions must necessarily change from face-to-face classrooms (Crippen, et al., 2013). As one teacher interviewed by Crippen, et al. (2013) put it, "I cannot see the development of conclusions. Only the conclusion itself. It is hard to be a part of the process when I basically only see a finished project; and can only respond to what they [students] have already concluded." (p. 1042). When combined with the constructivist view that the process is an important piece of the learning (Bruner, 1961), a crucial challenge of online science instruction is to find ways for the teacher to view and participate in students' process.

Jaber, et al. (2018) provide one possibility in their design of a responsive online professional development course for elementary and middle school teachers focusing on science practices. They began their course by asking participants whether gravity or air pressure drives the movement of water through a siphon. Participants were then challenged to collect and present evidence to help answer the question. In a discussion forum, participants posted photos of hands-on experiments they designed using materials available to them at home and shared the results of those experiments. As results of these experiments were shared on the forum, participants began developing and sharing possible explanations in the same discussion board. Participants began revising their explanations, conducting new experiments, and proposing new sub-questions in the forum as they engaged with their peers' ideas. This discussion forum made students' process visible to the instructors and allowed them to design instruction based on the ideas and questions posted to the forum.

Jang (2009) describes another approach utilized for a unit on calories in a blended secondary science classroom. The unit was introduced by asking students to work in groups on a web-based homework assignment to find methods of losing weight. Students then shared the results of their homework in a whole-class discussion. Next, students worked in groups to design experiments on the calorie content of different foods, the results of which were shared with the class. The lab was followed by a second online homework assignment that utilized open-ended questions building on concepts developed during the lab. Finally, students worked in groups to prepare an online presentation applying concepts from the unit. Both of the online homework assignments gave the instructor the opportunity to track students' conceptual development as they progressed through the unit. In addition, when presenting both the lab and the concluding project, students were eager to share the process they used to arrive at their results. While the teacher still relied on face-to-face instruction to intervene in students' process in real-time, the emphasis on sharing process, not just results, in the presentations gave the teacher opportunities to gain insight into the students' processes. In the context of an extended online course, this could enable the instructor to give feedback and coaching to encourage students' growth in the process of inquiry.

Several key characteristics were present in the courses examined by Jaber, et al. (2018) and Jang (2009). In both courses, instructors consciously fostered a culture that placed an explicit value on the process used by students. In addition, both courses provided clear avenues for students to share their process and receive feedback from both peers and instructors before arriving at a conclusion. This suggests that while it is challenging to make students' process visible to instructors in an online environment, it is not impossible.

**Conclusions**

Constructivist learning theories have an important influence on current reform movements towards inquiry-based science instruction. An important element of this instructional approach is implementing labs that are designed to meet specific instructional goals, provide opportunities for students to engage in sense-making and knowledge construction, and engage students in the process of inquiry. These labs should be integrated into the larger curriculum, rather than treated as independent experiences and part of a larger instructional approach rooted in responsive teaching that foregrounds student thinking and uses that thinking to inform instructional decisions.

While there are many who express a preference for hands-on labs, there is little empirical evidence that hands-on labs are inherently preferable to remote labs, virtual labs, or simulated labs. Depending on the goals of the instruction, each may serve a purpose within an inquiry-based science classroom. Regardless of the form a lab takes, the primary factor in whether it achieves constructivist goals is the way the lab is used in instruction. Therefore, in a purely online classroom the primary challenge is not in finding ways to conduct lab activities, but in finding ways for the teacher to observe and participate in students' process in order to be responsive to student thinking and guide students in developing their skills with the process of inquiry. Discussion forums and online assignments where students are explicitly asked to share their process can provide opportunities for teachers to gain insight into students' process in order to inform instructional decisions and guide students to develop the skills necessary for inquiry.

Implementing constructivist learning theories in online science classrooms is challenging, but not impossible. Instructors must carefully attend to the goals of instructional activities, including how lab experiences fit within the larger curriculum, take intentional steps to give students opportunities to share their process, and ensure instruction is responsive to student thinking and ideas. When an online course is designed with these considerations, it can be delivered in a way consistent with constructivist learning theories.